\documentstyle[aps,epsfig,multicol]{revtex}
\tightenlines
\begin{document}
\title{A New Coupling Potential for the Scattering of Deformed Light Heavy-Ions}
\author{I. Boztosun and W. D. M. Rae}
\address{Department of Nuclear Physics, University of Oxford, Keble Road OX1 3RH Oxford UK.}
\maketitle
\begin{abstract}
This letter introduces a new coupling potential to explain the
experimental data over wide energy ranges for a number of systems.
Within the coupled-channels formalism, this letter first shows the
limitations of the standard coupled-channels theory in the case
where one of the nuclei in the reaction is strongly deformed and
then, demonstrates that a global solution to the problems of light
heavy-ion reactions such as $^{12}$C+$^{12}$C, $^{16}$O+$^{28}$Si
and $^{12}$C+$^{24}$Mg can be found using a new second-derivative
coupling potential in the coupled-channels formalism. This new
approach consistently improves the agreement with the experimental
data for the elastic and inelastic scattering data as well as for
their excitation functions using constant or slightly
energy-dependent parameters.
\end{abstract}
\begin{multicols}{2}[]
\narrowtext
In this letter, building on a previous paper \cite{Boz1}, we
introduce a new type of coupling potential that explains the
scattering observables of the $^{16}$O+$^{28}$Si,
$^{12}$C+$^{24}$Mg and $^{12}$C+$^{12}$C systems in a consistent
way over wide energy ranges. These reactions have been extensively
studied and found to present many problems to which no consistent
solution has been provided yet using the standard coupled-channels
theory.

Over the last 40 years, a large body of experimental data has been
accumulated from the systematic studies of these reactions
\cite{Bra82,Sci97,Sto79}. Although the theoretical models, based
on dynamical models or purely phenomenological treatments, provide
reasonably good fits \cite{Bra82,Bra97,Kob84}, no unique model has
been proposed that explains consistently the elastic and inelastic
scattering data over wide energy ranges without applying any {\it
ad-hoc} procedures.

Consequently, the following problems continue to exist for light
heavy-ion reactions: $(1)$ explanation of anomalous large angle
scattering data; $(2)$ reproduction of the oscillatory structure
near the Coulomb barrier; $(3)$ the out-of-phase problem between
theoretical predictions and experimental data; $(4)$ simultaneous
fits of the individual angular distributions, resonances and
excitation functions (for the $^{12}$C+$^{12}$C system in
particular); $(5)$ the magnitude of the mutual-2$^{+}$ excited
state data in the $^{12}$C+$^{12}$C system is unaccounted for;
$(6)$ the deformation parameters ($\beta$ values): previous
calculations require $\beta$ values that are at variance with the
empirical values and are physically unjustifiable.

In the next section, we introduce the standard and new
coupled-channels models. Sections \ref{cc}, \ref{osi} and
\ref{mgc} are devoted to the application of the new coupling
potential to the analyzes of the $^{12}$C+$^{12}$C,
$^{16}$O+$^{28}$Si and $^{12}$C+$^{24}$Mg reactions. Finally, we
conclude in section \ref{diss}.
\section{The Model}
\vskip-0.5cm
The three systems we study ($^{16}$O+$^{28}$Si, $^{12}$C+$^{24}$Mg
and $^{12}$C+$^{12}$C) are quite different in many ways but they
share two common features: $(1)$ the elastic scattering data
suggest that there is weak absorption in the entrance channel in
each case and $(2)$ each reaction involves at least one nucleus
which is highly deformed.
\subsection{The Standard Coupled-Channels Calculations}
\vskip-0.25cm

In the standard coupled-channels calculations, we describe the
interaction between two nuclei with a deformed optical potential.
For all the reactions considered, the real potential is assumed to
have the square of a Woods-Saxon shape. The parameters of the real
potential for the  $^{16}$O+$^{28}$Si and $^{12}$C+$^{24}$Mg
systems are fixed as a function of energy and are not changed in
the present calculations. For the $^{12}$C+$^{12}$C system, it is
slightly energy-dependent. The parameters are shown in
table~\ref{table1}.

The imaginary part of the potential is the standard Woods-Saxon
shape. Only the depth or radius increased linearly with energy.

Since the target nuclei $^{28}$Si, $^{24}$Mg and $^{12}$C are
strongly deformed, it is essential to treat their collective
excitations explicitly in the framework of the coupled-channels
formalism. We assume that the target nuclei have static quadrupole
deformations, and that their excitations can be described in the
framework of the collective rotational model. The empirical values
for the deformation parameters ($\beta$), derived from the known
B(E2) values, are used in the present calculations.

Using this standard coupled-channels method, we found, as other
authors had  found, that it was not possible to find a consistent
solution over wide energy ranges to the above-mentioned problems
(see the discussions of results and figures below).
\subsection{New Coupling Potential}
\vskip-0.25cm
The limitations of the standard coupled-channels method, on the
one hand, and the resulting shape of the compound nucleus created
by the projectile and target nuclei, on the other hand, have
motivated us to use a second-derivative coupling potential. If we
consider two $^{12}$C nuclei approaching each other, the
double-folding model will generate an {\it oblate} potential which
is correct at large distances. When these two nuclei come close
enough, they create the compound nucleus $^{24}$Mg which is a {\it
prolate} nucleus in its ground state. In the standard method, the
folding model yields an oblate (attractive) potential for such a
configuration. It is not clear how well the double-folding model
describes a prolate nucleus with this oblate potential and this
may be the reason why the earlier calculations using a
double-folding model in the coupled-channels method were unable to
provide a consistent solution to the problems of such reactions.
In order to describe the above-mentioned configuration, the
coupling potential must be {\it oblate} (attractive) when two
$^{12}$C nuclei are at large distances and must be {\it prolate}
(repulsive) when they are at short distances \cite{Boz1}.
\begin{figure}[t]
\centering \epsfxsize=6.0cm \epsfbox{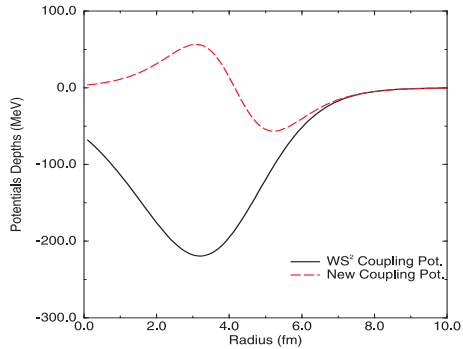} \vskip+0.25cm
\caption{For $^{16}$O+$^{28}$Si, the comparison of the {\it
standard coupling potential} which is the first derivative of the
central potential and our {\it new coupling potential} which is
parameterized as the 2$^{nd}$ derivative of Woods-Saxon shape.}
\label{fig:16Ocoup}
\end{figure}
The standard and new coupling potentials are compared in figure
\ref{fig:16Ocoup}. In the new calculations, we employed the same
method with the same potentials. The parameters are shown in
table~\ref{table1}. The empirical deformation parameter
($\beta_{2}$) is used in these calculations.

\section{$^{12}$C+$^{12}$C}
\label{cc} \vskip-0.5cm
The first system we consider is the $^{12}$C+$^{12}$C, which has
been studied extensively over the last 40 years. Our analysis
consists of a {\it simultaneous} investigation of the elastic
scattering, single-2$^{+}$ and mutual-2$^{+}$ excitation inelastic
scattering data from $E_{Lab}$=20.0 MeV to 126.7 MeV. In such a
large energy range, we also consider the 90$^{\circ}$  elastic and
inelastic excitation functions.

The conventional folding model potentials fail to reproduce
certain aspects of the data such as the reproduction of the gross
structure in the 90$^{\circ}$ elastic scattering excitation
function and a simultaneous consistent description of the elastic,
single-2$^{+}$ and mutual-2$^{+}$ states data. So far, no model
has been able to predict the magnitude of the mutual-2$^{+}$ state
data correctly over a wide energy range. The standard
coupled-channels model using double-folding potentials
underestimates its magnitude by a factor of 3 to 10 and the
single-2$^{+}$ state results are too oscillatory with respect to
experimental data~\cite{Sto79,Fry97,Rae97,Sak99,Wol82}. These
problems have remained unsolved so far. Clearly, the
$^{12}$C+$^{12}$C system has numerous problems to which no
consistent global solution has been provided yet.

\begin{figure}[t]
\epsfxsize=8.2cm \centering \epsfbox{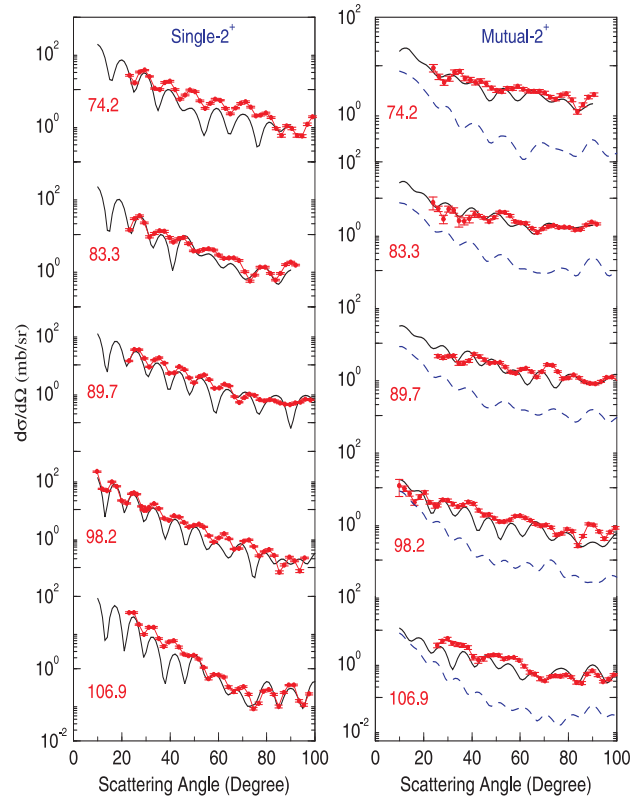} \vskip+0.25cm
\caption{The $^{12}$C+$^{12}$C system: The results of the single
and mutual-2$^{+}$ states. The {\it solid} lines are the results
of {\it the new coupling potential}, while the {\it dashed} lines
are the results of {\it standard coupled-channels model}.}
\label{comp}
\end{figure}

We have also observed such results in our standard
coupled-channels calculations, as shown only for the
mutual-2$^{+}$ case in figure \ref{comp} with dashed lines.
Varying the parameters and changing the shape of the real and
imaginary potentials and some other attempts do not provide a
complete solution to the problems of this reaction~\cite{Boz1}.

Using the new coupling potential, we have been able to obtain
excellent agreement with all the available experimental data for
the ground, single-2$^{+}$ and mutual-2$^{+}$ states as well as
the 90$^{\circ}$ elastic scattering excitation function. This new
approach also solves the magnitude problem of the mutual-2$^{+}$
excitation inelastic scattering data, which has been outstanding
problem of this reaction. Some of the results for the
single-2$^{+}$ and mutual-2$^{+}$ states are shown in figure
\ref{comp} in comparison with the standard ones. All results can
be found in ref.~\cite{Boz1}.

\section{$^{16}$O+$^{28}$S\lowercase{i}}
\label{osi} \vskip-0.5cm

The second system we consider is $^{16}$O + $^{28}$Si, which shows
anomalous large angle scattering (ALAS). In the present work, we
consider an extensive {\it simultaneous} investigation of the
elastic and inelastic scattering of this system at numerous
energies from $E_{Lab}$=29.0 to 142.5 MeV. In this wide energy
range, the excitation functions for the ground and single-2$^{+}$
states are also analyzed \cite{Boz2}.

Several {\it ad-hoc} models have been proposed to explain the
experimental data, but no satisfactory microscopic models have
been put forward yet. The most satisfactory explanation proposed
so far is that of Kobos and Satchler \cite{Kob84} who attempted to
fit {\it only} the elastic scattering data with a microscopic
double-folding potential. However, these authors had to use some
small additional {\it ad-hoc} potentials, which create a deepening
in the surface region of the potential, to obtain a good agreement
with the experimental data. Without the additional small
potentials, they could not reproduce the experimental data. We
have shown that this deepening of the real potential in the
surface region takes into account the coupling effect in an {\it
ad-hoc} way~\cite{Boz4}.
\begin{figure}
\epsfxsize=6.25cm \centering \epsfbox{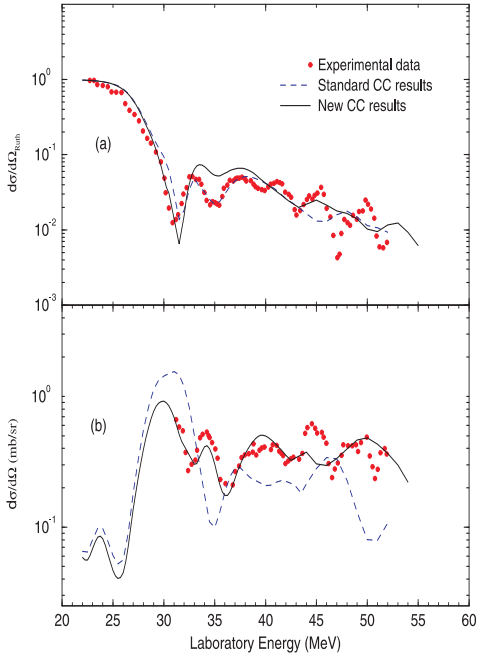} \vskip+0.25cm
\caption{The $^{16}$O+$^{28}$Si system: 180$^{\circ}$ excitation
function for $(a)$ ground and $(b)$ 2$^{+}$ states respectively.
The {\it dashed} lines are the results of {\it standard}
coupled-channels calculations and {\it solid} lines are the
results obtained using new coupling potential.} \label{16Oexccomp}
\end{figure}
Using the standard coupled-channels method, we obtained a very
good agreement with the elastic scattering data. However, for the
single-2$^{+}$ state data, the theoretical predictions were too
oscillatory and in particular, towards large angles at {\it
higher} energies, the out-of-phase problem between the theoretical
predictions and the experimental data is observed. This problem is
clearly seen in the 180$^{\circ}$ excitation function results,
shown in figure \ref{16Oexccomp}-b. A number of models have been
proposed, ranging from isolated resonances to cluster exchange
between the projectile and target nucleus, to solve these problems
(see ref. \cite{Bra82} for a detailed discussion).

We have also attempted to overcome these problems by considering:
$(i)$ changes in the real and imaginary potentials, $(ii)$
inclusion of 6$^{+}$ excited state, $(iii)$ changes in the
$\beta_{2}$ value and some other attempts failed to solve the
problems~\cite{Boz2}. We were unable to get an agreement with the
elastic and the 2$^{+}$ inelastic data as well as the
180$^{\circ}$ excitation functions simultaneously within the
standard coupled-channels formalism.

However, as shown in figure~\ref{16Oexccomp}, the new coupling
potential has solved the out-of-phase problem for the
180$^{\circ}$ excitation functions and fits the ground state and
2$^{+}$ state data simultaneously. To our knowledge, this has not
been achieved over such a wide energy range (see ref.~\cite{Boz2}
for all the results).

\section{$^{12}$C+$^{24}$M\lowercase{g}}
\label{mgc} \vskip-0.5cm

The final example we consider is the $^{12}$C + $^{24}$Mg system.
The angular distributions oscillate strongly near the Coulomb
barrier and the data manifest ALAS. Our analysis consists of a
{\it simultaneous} investigation of the elastic and inelastic
scattering data at numerous energies from $E_{Lab}$=16.0 to 24.0
MeV~\cite{Boz3}.

The most extensive study for this system was carried out by Sciani
{\it et al} \cite{Sci97}. The authors could only fit these data
with $Q$-value dependent potential parameters in a rather {\it
ad-hoc} fashion. Without $Q$-dependent potentials, they observed
that the theoretical predictions and the experimental data for the
elastic and inelastic scattering data were completely
out-of-phase.
\begin{figure}
\epsfxsize=8.0cm \centering \epsfbox{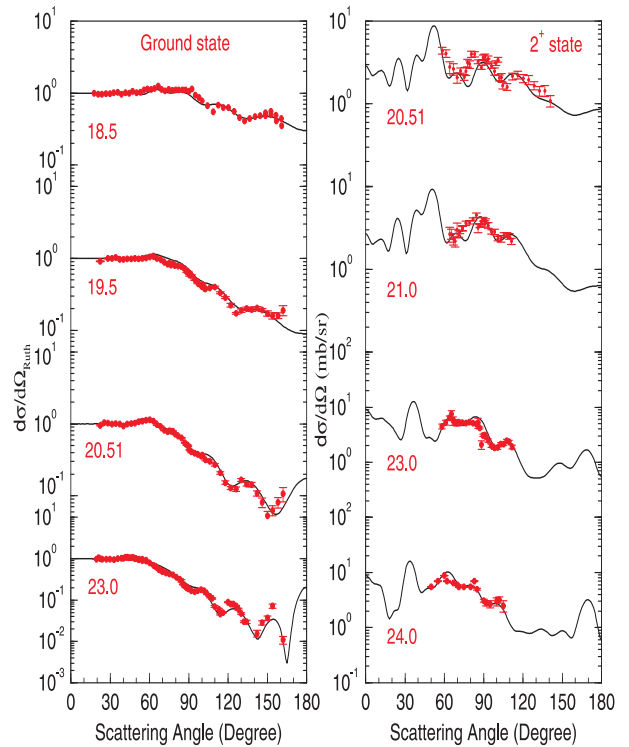} \vskip+0.25cm
\caption{The $^{12}$C+$^{24}$Mg system: The ground and 2$^{+}$
states results, obtained using the new coupling potential.}
\label{cmg}
\end{figure}
In the present calculations, we have studied the data of ref.
\cite{Sci97} as well as some inelastic scattering data of
ref.~\cite{Car76}. In the standard coupled-channels calculations,
we also found the out-of-phase problem which was in conformity
with the findings of Sciani {\it et al}. However, using our new
coupling potential, we obtained excellent agreement with the
experimental data. Some of the individual angular distribution
fits for the ground and 2$^{+}$ states are shown in figure
\ref{cmg} (see ref.~\cite{Boz3} for all the results).
\section{Discussion \lowercase{and} Summary}
\label{diss} \vskip-0.5cm
The effect of this new coupling potential on the scattering has
been probed through the total inverted potential, {\it i.e.} the
dynamical polarization potential (DPP) plus the bare potential,
obtained by inversion of the S-matrix. The most effective and
sensitive regions of the new coupling potential have been
identified. We have precisely determined the effect of the
inclusion of the excited states on both the real and imaginary
potentials as well as the effect of the new coupling
potential~\cite{Boz4}.

\begin{figure}
\epsfxsize=6.25cm  \centering \epsfbox{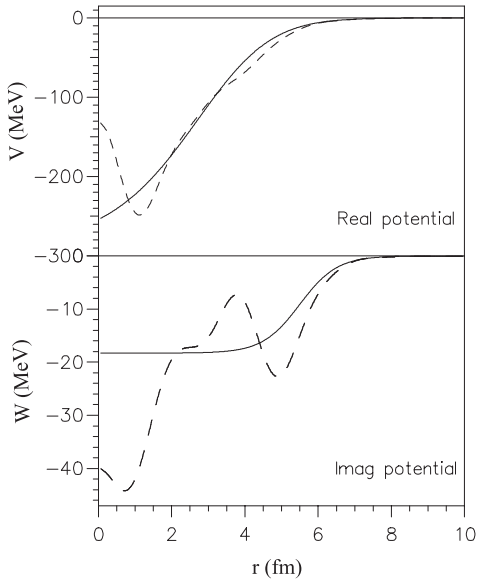} \vskip+0.25cm
\caption{The comparison of the bare potentials (solid lines) and
the total inverted potentials (dashed lines), found by inverting
the S-matrix of the new coupled-channels calculations at
$E_{Lab}$=126.7 MeV for the $^{12}$C+$^{12}$C system.} \label{dpp}
\end{figure}

We have observed that for the standard coupling potential case,
the added attraction is almost the same as the added absorption.
On the other hand, for the new coupling potential case, the added
attraction is much greater than the added absorption. Therefore,
as shown in figure \ref{dpp}, the new coupling potential creates a
deepening in the surface region of the real potential~\cite{Boz4}.
Thus, it increases the interaction radius.

This deepening can be related to the superdeformed structures of
the compound nuclei~\cite{Boz1}. It is well known that the light
heavy-ion reactions are very sensitive to the surface region of
the potential and this deepening created by the new coupling
potential clearly makes a major improvement in explaining the
experimental data over such large energy ranges for different
reactions. This deepening has been included by many authors in an
artificial way in their calculations and they managed to explain
the elastic scattering data successfully (see
~\cite{Kob84,Mac82,Ord86} and references therein)

To summarize, while these three systems show quite different
properties and problems, a unique solution has come from a new
coupling potential. The importance of the new approach should be
underlined here because it does not only fit the present
experimental data, but it also leads to other novel and testable
predictions \cite{Boz1,Bre01}. To our knowledge, this has not been
yet achieved over such wide energy ranges for many different
reactions simultaneously. Our work reveals that there is no reason
for the coupling potential to have the same energy dependence as
the central term and studies using this new coupling potential may
lead to new insights into the formalism and also a new
interpretation of such reactions. Therefore, this work represents
an important step towards a new understanding of the elastic and
inelastic scattering of deformed light heavy-ion reactions.

\begin{table}
 \caption{The parameters of the central and coupling potentials for
the reactions studied.}
\begin{center}
\begin{tabular}{||l|lll||lll||}
    &   Central  &   &   &   Coupling   &   &    \\     \hline
&   V    &  $r_{0}$ &   a   &   V   &   $r_{0}$ &   a   \\  \hline
$^{16}$O+$^{28}$Si & 750.5 & 0.749 & 1.4 & 155.0 & 0.748 & 0.81 \\
$^{12}$C+$^{24}$Mg & 427.0 & 0.865 & 1.187 & 185.0 & 0.710 & 0.62 \\
$^{12}$C+$^{12}$C&$\sim$290.0&$\sim$0.8&$\sim$1.28&210.0&$\sim$0.67&$\sim$0.68
 \label{table1}
\end{tabular}
\end{center}
\end{table}
\end{multicols}
\end{document}